\documentclass[twocolumn,preprintnumbers,amsmath,amssymb,superscriptaddress,pre,showpacs]{revtex4}
\usepackage[dvips]{graphicx}
\usepackage{dcolumn}
\usepackage{bm}
\usepackage[dvips]{epsfig}
\newcommand{\gp}{\dot{\gamma}}
\newcommand{\dd}{\rm d} 

\begin{document}

\title{From stress-induced fluidization processes\\ to Herschel-Bulkley behaviour in simple yield stress fluids}

\author{Thibaut Divoux}
\affiliation{Universit\'e de Lyon, Laboratoire de Physique, \'Ecole Normale Sup\'erieure de
Lyon, CNRS UMR 5672, 46 All\'ee d'Italie, 69364 Lyon cedex 07, France.}
\author{Catherine Barentin}
\affiliation{Laboratoire de Physique de la Mati\`ere Condens\'ee et Nanostructures, Universit\'e de Lyon; Universit\'e Claude Bernard Lyon I, CNRS UMR 5586 - 43 Boulevard du 11 Novembre 1918, 69622 Villeurbanne cedex, France.}
\author{S\'ebastien Manneville}
\affiliation{Universit\'e de Lyon, Laboratoire de Physique, \'Ecole Normale Sup\'erieure de
Lyon, CNRS UMR 5672, 46 All\'ee d'Italie, 69364 Lyon cedex 07, France.}
\affiliation{Institut Universitaire de France.}
\date{\today}

\begin{abstract}
Stress-induced fluidization of a simple yield stress fluid, namely a carbopol microgel, is addressed through extensive rheological measurements coupled to simultaneous temporally and spatially resolved velocimetry. These combined measurements allow us to rule out any bulk fracture-like scenario during the fluidization process such as that suggested in [Caton {\it et al., Rheol Acta}, 2008, {\bf 47}, 601-607]. On the contrary, we observe that the transient regime from solidlike to liquidlike behaviour under a constant shear stress $\sigma$ successively involves creep deformation, total wall slip, and shear banding before a homogeneous steady state is reached. Interestingly, the total duration $\tau_f$ of this fluidization process scales as $\tau_f \propto 1/(\sigma - \sigma_c)^{\beta}$, where $\sigma_c$ stands for the yield stress of the microgel, and $\beta$ is an exponent which only depends on the microgel properties and not on the gap width or on the boundary conditions. Together with recent experiments under imposed shear rate [Divoux {\it et al., Phys. Rev. Lett.}, 2010, {\bf 104}, 208301], this scaling law suggests a route to rationalize the phenomenological Herschel-Bulkley (HB) power-law classically used to describe the steady-state rheology of simple yield stress fluids. In particular, we show that the {\it steady-state} HB exponent appears as the ratio of the two fluidization exponents extracted separately from the {\it transient} fluidization processes respectively under controlled shear rate and under controlled shear stress.
\end{abstract}
\maketitle

\section{Introduction}

Yield stress fluids (YSF) are widely involved in manufactured products such as creams, gels, or shampoos. These materials are characterized by a transition from solidlike to liquidlike above the yield stress $\sigma_c$, which is of primary importance at both the manufacturing stage and the end-user level \cite{Coussot:2006}. Recently it was recognized that simple YSF, which mainly consist in emulsions, foams, and carbopol microgels, can be clearly distinguished from thixotropic YSF \cite{Moller:2009b}: in {\it steady state} the former ones can flow homogeneously at vanishingly small shear rates under controlled stress \cite{Coussot:2009,Ovarlez:2010} while the latter exhibit a finite critical shear rate \cite{Becu:2006,Ragouilliaux:2007}.
Still, in spite of its importance for applications, the {\it transient} fluidization process of simple YSF has remained largely unexplored and previous works have focused either on global rheometry under an applied stress \cite{Uhlherr:2005,Coussot:2006,Caton:2008} or on time-resolved local velocimetry under controlled shear rate \cite{Hohler:2005,Becu:2005,Divoux:2010}. Thus, detailed local information concerning the fluidization of a simple YSF {\it under applied shear stress} are still lacking, which prevents to make clear connections with observations under imposed shear rate and with steady-state rheology.

In this article we report a temporally and spatially resolved study of the stress-induced fluidization of carbopol microgels through ultrasonic echography. Our aim is to address the following basic questions: (i) What is the fluidization scenario of such a simple YSF under imposed shear stress? (ii) How does it compare to imposed shear rate experiments? (iii) Can one make a connection between these transient fluidization processes and the steady-state rheology, which is well described by the Herschel-Bulkley (HB) law \cite{Hohler:2005,Mason:1996a,Roberts:2001,Piau:2007,Divoux:2010}? Here, we show using an ultrasonic velocimetry technique that carbopol microgels submitted to a constant shear stress $\sigma$ under rough boundary conditions successively exhibit creep deformation, total wall slip, and shear banding before reaching a homogeneous steady state. A close inspection of the backscattered ultrasonic signals allows us to rule out a scenario involving bulk fracture of the 
 material. The duration of the  fluidization process decreases as a power-law with the reduced shear stress $\sigma-\sigma_c$. This power law only depends on the sample preparation protocol and not on boundary conditions or on the cell gap. Together with recent experiments under imposed shear rate \cite{Divoux:2010}, this provides for the first time a direct link between the yielding dynamics of a simple YSF and the HB law which accounts for its steady-state rheology.

\begin{figure*}[!t]\tt
\centering
\includegraphics[width=1.1\columnwidth]{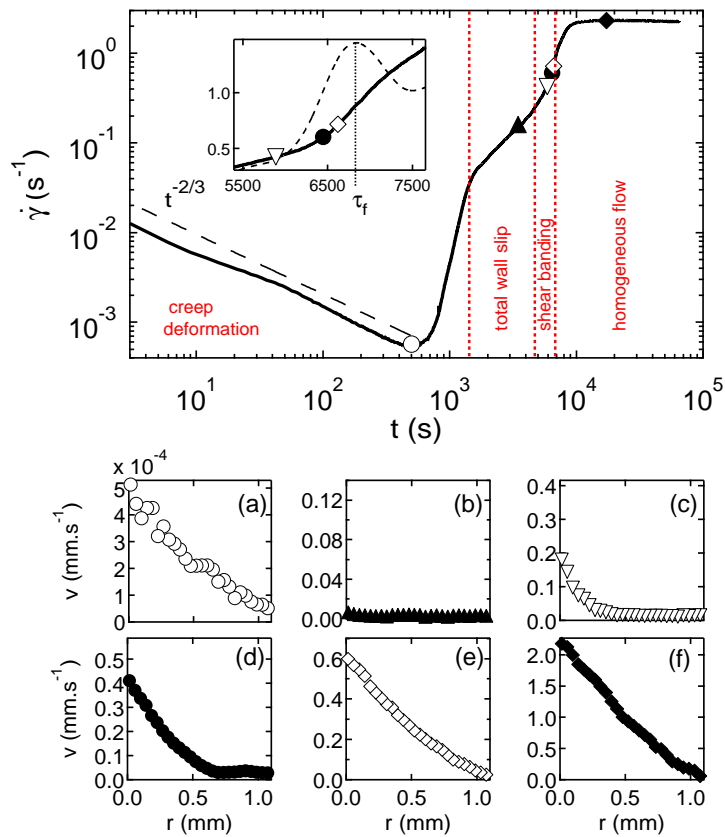}
\caption{Top: Shear-rate $\gp$ vs time $t$ for a shear stress $\sigma=37$~Pa applied at $t=0$ (batch 1, rough BC). Inset: zoom on the shear-banding regime. The dashed curve shows $\dd\gp(t)/\dd t$ whose maximum corresponds to $\tau_f$. Bottom: velocity profiles $v(r)$, where $r$ is the distance to the rotor, at different times [symbol, time (s)]: ($\bigcirc$, 500), ($\blacktriangle $, 3460), ($\triangledown $, 5890), ($\bullet $, 6450), ($\diamond$, 6620), ($\blacklozenge$, 17349). On each profile, the upper value of the velocity scale is set to the current rotor velocity $v_0$ so that the slip velocity $v_s$ at the rotor can be read directly as $v_s=v_0-v(r=0)$.}
\label{fig.1}
\end{figure*}

\section{Experimental}

\subsection{Sample preparation}

\subsubsection{Working system: carbopol microgels.~~}

Our working system is a microgel made of carbopol ETD~2050 which comprises homo- and copolymers of acrylic acid highly cross-linked with a polyalkenyl polyether \cite{Roberts:2001,Baudonnet:2004}. The microgel is traditionally prepared in two steps: (i) the polymer is dispersed in water leading to a suspension of carbopol aggregates and (ii) a neutralizing agent is added (in our case sodium hydroxyde) leading to polymer swelling and to microgel formation. The microstructure of such microgels consists in an assembly of soft jammed swollen polymer particles, with typical size ranging roughly from a few microns to roughly 20 microns\cite{Ketz:1988,Kim:2003,Oppong:2006} depending on the type of carbopol \cite{Baudonnet:2004}, its concentration \cite{Roberts:2001}, the final value of the pH \cite{Lee:2011}, the type of neutralizing agent \cite{Kim:2003} and, last but not least, the stirring speed during neutralization \cite{Baudonnet:2004}. Carbopol microgels exhibit good temperat
 ure stability \cite{Islam:2004}. They are also known in the literature to be non-aging, non-thixotropic {\it simple} YSF \cite{Piau:2007,Coussot:2009,Moller:2009a,Benmouffok:2010} and their steady-state flow curve nicely follows the HB law: 
\begin{equation}
\sigma=\sigma_c+\tilde{\eta}\,\gp^n\,,
\label{eq1}
\end{equation}
where $\gp$ is the shear rate and $n=0.3$--0.6 depending on the type of carbopol and its concentration \cite{Divoux:2010,Roberts:2001,Oppong:2006,Coussot:2009}.

\subsubsection{Sample preparation protocol.~~}

For our study we prepare two kinds of samples: traditional samples on the one hand, and samples that are seeded with micronsized glass spheres on the other hand, in order to use ultrasonic speckle velocimetry (USV) \cite{Manneville:2004a} simultaneously to standard rheological measurements.
The detailed protocol for seeded samples is as follows: we first add 0.5\%~wt. of hollow glass spheres (Potters, Sphericel, mean diameter 6~$\mu$m, density 1.1) in ultrapure water; pH increases roughly from 7 to 8. As carbopol is hydrosoluble only for pH$<7$, we add one or two drops of concentrated sulfuric acid (H$_2$SO$_4$, 96~\%) to make the pH drop to roughly 5. The glass sphere suspension is heated at $50^{\circ}$C and the carbopol powder is carefully dispersed under magnetic stirring at 300~rpm for 40~min, at a weight fraction $C$, with $C$ ranging from 0.5 to 3\%~wt. The mixture is then left at rest at room temperature for another 30~min, after which pH~$\simeq 3$. Finally we neutralize the solution with sodium hydroxide (NaOH, concentration 10 mol.L$^{-1}$) until pH~$=7.0\pm0.5$ while stirring manually. This leads to a carbopol microgel which is finally centrifuged for 10~min at 2500~rpm to get rid of trapped bubbles. As for traditional samples without glass beads, the protocol starts directly by adding the carbopol powder to a heated volume of ultrapure water and continues as explained above. 

\subsubsection{Influence of the preparation protocol on the batch properties.~~}

We emphasize that the final microgel macroscopic properties are quite sensitive to the preparation protocol. In particular, it is well known in the literature that during the neutralization step, (i)~the way the base is added (drop by drop or all at once) as well as (ii)~the exact final value of the pH and (iii)~the stirring speed during the neutralization process influence the values of the parameters of the HB model \cite{Curran:2002,Baudonnet:2004}. This derives from the fact that these three parameters control the particle size distribution of the microgel \cite{Lee:2011,Baudonnet:2004}. In this paper, we take good care to neutralize our samples in a reproducible fashion. Nonetheless, from batch to batch, the final pH value of the microgel varies in the range $6.5<$~pH~$<7.5$. Therefore, when comparing different geometries, gaps, or boundary conditions, we pay special attention to use results obtained on a single batch, so that the preparation protocol does not introduce any bias. We will mainly report data obtained on two different batches of carbopol weight fraction $C=1$~\% wt. and seeded with glass spheres, noted batch 1 and batch 2, prepared separately but following the same protocol. We will also discuss the influence of the carbopol concentration $C$ on four different traditional unseeded batches prepared separately: $C=$~0.5, 0.7, 1, and 3 \% wt. 

\subsubsection{Influence of the seeding glass spheres.~~}

Linear viscoelasticity measurements show that the addition of hollow glass spheres generally slightly stiffens the system (by at most 10\%)\cite{Divoux:2011}. However, we shall check throughout the whole manuscript that traditional samples and seeded samples exhibit the exact same rheological trends, which demonstrates that the acoustic contrast agents play no significant role in the fluidization scenario under imposed shear stress.
 
\subsection{Experimental setup and protocol}

\subsubsection{Rheological setup.~~}

Rheological measurements are performed with a stress-controlled rheometer (Anton Paar MCR301). Two different small-gap Couette cells were used to test the influence of the boundary conditions (BC) on the fluidization process: a rough cell (surface roughness $\delta\simeq 60~\mu$m obtained by gluing sandpaper to both walls, rotating inner cylinder radius $R_{\rm int}=23.9$~mm, gap width $e=1.1$~mm, and height $h=28$~mm) and a smooth Plexiglas cell [$\delta\simeq 15$~nm (AFM measurements), $R_{\rm int}=24$~mm, $e=1$~mm, and $h=28$~mm]. Also, to test the influence of both the geometry and the gap, experiments were performed with a plate-plate geometry (radius 21 mm, gap width $e=1$ and $e=3$~mm) with two different boundary conditions: rough (glued sandpaper, $\delta\simeq 46$~$\mu$m) and smooth [glass, $\delta\simeq 6$~nm (AFM measurements)]. Finally, note that for both geometries, we use a solvent trap including a cover and a small water tank so as to efficiently prevent evaporation.

\subsubsection{Local velocity measurements.~~}

Velocity profiles are measured at about 15~mm from the cell bottom through ultrasonic speckle velocimetry (USV) as described in details by Manneville {\it et al.} \cite{Manneville:2004a} In brief, USV relies on the analysis of ultrasonic speckle signals that result from the interferences of the backscattered echoes of successive incident pulses of central frequency 36MHz generated by a high-frequency piezo-polymer transducer (Panametrics PI50-2) connected to a broadband pulser-receiver (Panametrics 5900PR with 200 MHz bandwidth). The speckle signals are sent to a high-speed digitizer (Acqiris DP235 with 500 MHz sampling frequency) and stored on a PC for postprocessing. A cross-correlation algorithm yields the local displacement from one pulse to another as a function of the radial position across the gap with a spatial resolution of 40~$\mu$m. After a calibration step using a Newtonian fluid, tangential velocity profiles are then obtained by averaging over 10 to 1000 successive cross-correlations depending on the desired temporal resolution. During the creep experiments considered here, the shear rate $\gp$ strongly varies in time so that the technique described in \cite{Manneville:2004a} for a fixed shear rate has to be slightly modified: we use the analog output of the MCR 301 rheometer to monitor the rotor velocity in real-time and to constantly adapt the repetition frequency of the ultrasonic pulses to the current $\gp(t)$ in order to follow the dynamics.

\subsubsection{Experimental protocol.~~}

Before starting an experiment on a fresh sample, a strong preshear is applied for 1 min at +1000~s$^{-1}$ and for 1 min at -1000~s$^{-1}$ to erase the loading history. The viscoelastic moduli are then monitored for 2~min. We found that both the elastic and the viscous moduli do not vary significantly after 2 min. Finally, the sample is left at rest for 1 min to ensure that a reproducible initial state is reached. Let us note that we used the exact same protocol for experiments performed under controlled shear rate \cite{Divoux:2010} so as to compare results obtained for both applied shear stress and applied shear rate.

\begin{figure}[t]\tt
\centering
\includegraphics[width=0.9\linewidth]{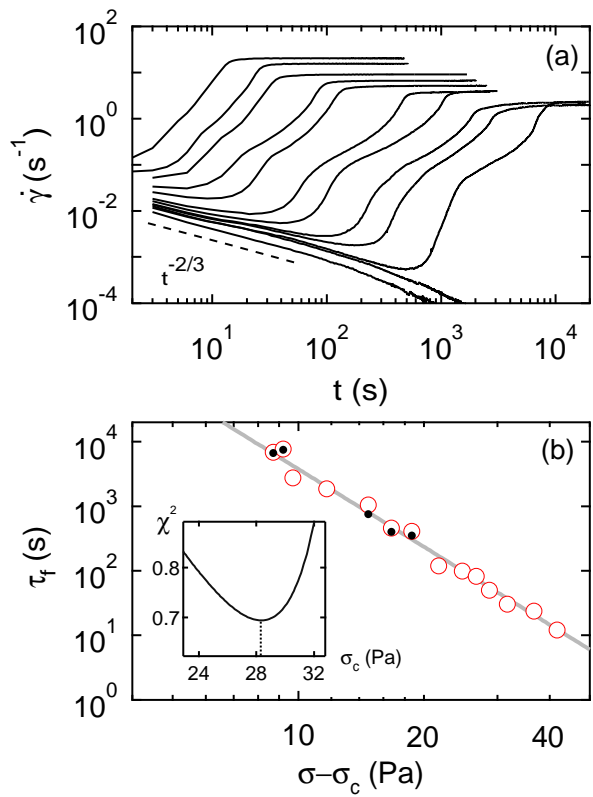}
\caption{(a) Shear rate $\gp$ vs time for $\sigma=35$, 36, 37, 38, 40, 43, 45, 50, 55, 60, 65, and 70~Pa (from bottom to top) for batch 1 and rough BC. (b) Fluidization time $\tau_f$ extracted from the second inflection point of $\gp(t)$ ($\circ$) and from USV ($\bullet$) vs the reduced shear stress $\sigma-\sigma_c$ with $\sigma_c=28.3$~Pa. The gray line corresponds to the best fit $\tau_f=B/(\sigma-\sigma_c)^{\beta}$ with $B=(3.77\pm 0.07)\cdot 10^7$ and $\beta=4.0 \pm 0.1$. The inset shows the least square determination of $\sigma_c$ (see text).}
\label{fig.2}
\end{figure}

\section{Results}

 \subsection{General scenario under rough boundary conditions}

Velocity profiles are measured simultaneously to the global shear rate $\gp(t)$ during long creep experiments under an imposed shear stress $\sigma$. As shown in Fig.~\ref{fig.1} for $\sigma=37$~Pa and rough BC, the shear rate follows an S-shaped curve with an initial decay characterized by a power law $\dot \gamma (t) \propto t^{-2/3}$, which is strongly reminicent of the Andrade creep observed in plastically deforming crystals \cite{Miguel:2002,Nechad:2005} and some viscoelestic materials \cite{Bauer:2006}. This slow decay is followed by a strong increase of $\gp(t)$ before a steady-state is reached after about $10^4$~s. Local velocimetry allows us to further distinguish four regimes in creep experiments. The microgel first experiences {\it creep deformation} for $t\lesssim 1500$~s with extremely small velocities ($<1~\mu$m.s$^{-1}$) [Fig.~\ref{fig.1}(a)]. At $t\simeq 1500$~s, the microgel fails at the inner wall and, in spite of rough BC, it enters a regime of {\it total wall slip}  which lasts until $t\simeq 4700$~s [Fig.~\ref{fig.1}(b)]. For  $4700\lesssim t\lesssim 6600$~s, a {\it shear band} nucleates at the rotor and grows across the gap [Fig.~\ref{fig.1}(c)-(d)]. Full fluidization of the sample is achieved at $t=\tau_f\simeq 6600$~s, after which the velocity profiles remain linear while the slip velocity vanishes [Fig.~\ref{fig.1}(e)-(f)]. Interestingly, we observe that complete fluidization coincides almost exactly with the second inflection point of the curve $\gp(t)$ [inset of Fig.~\ref{fig.1}] so that the fluidization time $\tau_f$ may also be defined as the time at which  $\dd\gp(t)/\dd t$ passes through a maximum. As seen in Fig.~\ref{fig.2}(b) and Fig.~\ref{fig.3}(b), these two independent estimations of $\tau_f$ lead to negligible discrepancies.

 \subsection{Influence of the boundary conditions}
 \label{influenceBC}

\begin{figure}[t]\tt
\centering
\includegraphics[width=0.95\linewidth]{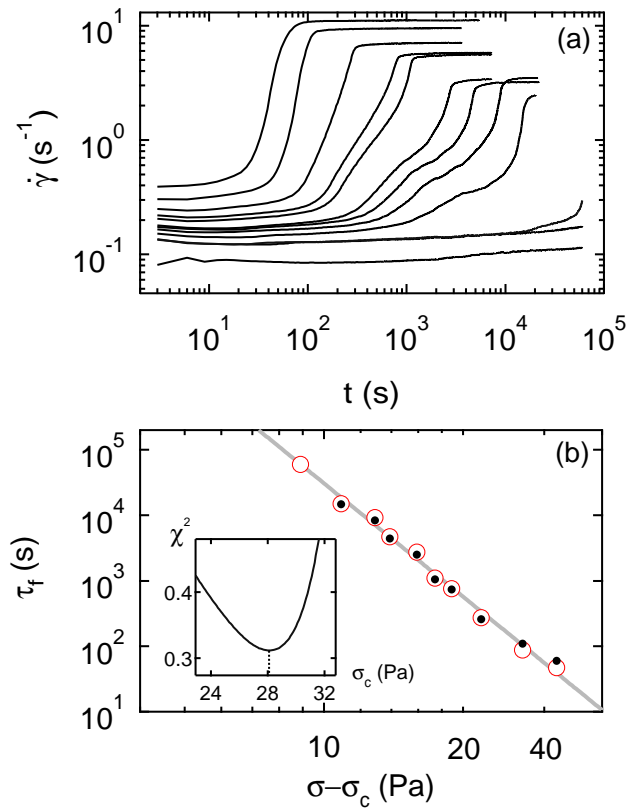}
\caption{(a) Shear rate $\gp$ vs time for $\sigma=32$, 36, 37, 39, 41, 42, 44, 45.5, 47, 50, 55, and 60~Pa (from bottom to top) for batch 2 and smooth BC. (b) Fluidization time $\tau_f$ extracted from the second inflection point of $\gp(t)$ ($\circ$) and from USV ($\bullet$) vs the reduced shear stress $\sigma-\sigma_c$ with $\sigma_c=28.1$~Pa. The gray line corresponds to the best fit $\tau_f=B/(\sigma-\sigma_c)^{\beta}$ with $B=(1.74\pm 0.04)\cdot 10^{10}$ and $\beta=5.75\pm 0.15$. The inset shows the least square determination of $\sigma_c$ (see text).}
\label{fig.3}
\end{figure}

As reported in Fig.~\ref{fig.2} and Fig.~\ref{fig.3}, creep experiments were repeated for $\sigma$ ranging from 35 to 70 Pa (32 to 60 Pa resp.) for rough (smooth resp.) BC . The above fluidization scenario is very robust and also holds for smooth BC. In this last case however, no creep deformation is observed and the slip regime starts almost instantly. Correspondingly $\gp(t)$ does not show any power-law behaviour at short times [Fig.~\ref{fig.3}(a)]. Nonetheless the rest of the process is identical to the one described for rough BC. Moreover, as seen in Fig.~\ref{fig.2}(b) and Fig.~\ref{fig.3}(b), the duration of the transient regime sharply decreases as the stress is increased. We found that the best way to fit the $\tau_f$ vs $\sigma$ data is a power law of the shear stress reduced by some well-defined critical stress $\sigma_c$: 
\begin{equation}
\tau_f = B/(\sigma-\sigma_c)^{\beta}.
\end{equation}
The value of $\sigma_c$ chosen here is the one that minimizes the $\chi^2$ of linear fits of $\ln \tau_f$ vs $\ln(\sigma-\sigma_c)$ [see insets of Fig.~\ref{fig.2}(b) and Fig.~\ref{fig.3}(b)]. Since the sample will {\it never} fluidize for $\sigma<\sigma_c$, this critical stress can be interpreted as the yield stress of the material. At this stage, it is important to note that our least-square procedure for measuring $\sigma_c$ from the fluidization time completely differs from any standard way of estimating the (static) yield stress under controlled shear stress or from short-time stress overshoots under controlled shear rate \cite{Divoux:2011}. We shall show below that this $\sigma_c$ indeed nicely coincides with the (dynamic) yield stress inferred from steady-state flow curves, in the limit of vanishing shear rates. Moreover, as discussed in more details in section~\ref{InfluenceGeoConc}, the exponents found in Fig.~\ref{fig.2}(b) and Fig.~\ref{fig.3}(b), $\beta=4.0$ and 5.75 for batch 1 and 2 respectively, only depend on the batch and not on the gap width or on the boundary conditions. We shall therefore conclude that the fluidization exponent is only function of the sample preparation, most probably through the final value of the pH of the microgel, as already observed under imposed shear rate \cite{Divoux:2010}.

\begin{table}[b]
\small
  \caption{\ Evolution of the fluidization exponent $\beta$ and prefactor $B$ with the carbopol concentration $C$.}
  \label{tab.1}
  \begin{tabular*}{0.5\textwidth}{@{\extracolsep{\fill}}lllll}
    \hline
     $C$ (\% wt.) & $B$ (Pa$^{\beta}$.s) & $\beta$ \\
    \hline
    0.5 &  1.48$\cdot$10$^5$	& $3.4 \pm 0.1$\\
    0.7 &  17.44$\cdot$10$^5$ 	& $3.75 \pm 0.10$\\
    1 &   5.86$\cdot$10$^{13}$ 	& $7,0\pm 0.6$\\
    3 &  1.19$\cdot$10$^{21}$	& $8.0 \pm$ 0.3\\
    \hline
  \end{tabular*}
\end{table}

 \subsection{Influence of the geometry and carbopol concentration}
 \label{InfluenceGeoConc}

\begin{figure}[t]\tt
\centering
\includegraphics[width=0.9\linewidth]{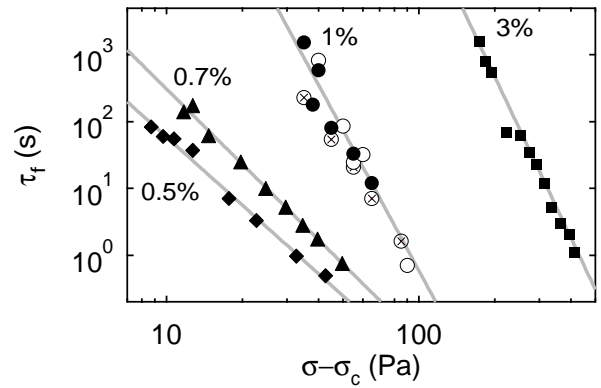}
\caption{Fluidization time $\tau_f$ extracted from the second inflection point of $\gp(t)$ vs the reduced shear stress $\sigma-\sigma_c$ for various carbopol concentrations in a rough plate-plate geometry of gap 1 mm: [symbol, \% wt., $\sigma_c$ (Pa)]: ($\blacklozenge$, 0.5, 17.3); ($\blacktriangle$, 0.7, 30.3); ($\bullet$, 1, 30); ($\blacksquare$, 3, 56.5). For the same batch of concentration 1 \% wt., we test the influence of the boundary conditions and of the gap width: (symbol, gap, BC): ($\bullet$, 1~mm, rough); ($\circ$, 3~mm, rough); ($\otimes $, 1~mm, smooth). Here, smooth BC correspond to glass plates.}
\label{fig.4}
\end{figure}

Creep experiments were performed in both rough and smooth plate-plate geometries for various mass concentrations: $C=0.5, 0.7, 1$ and 3 \% wt. In this case, the samples are free of seeding glass spheres. The shear rate responses $\gp(t)$ (not shown) present the same characteristics as those of Fig.~\ref{fig.2}(a) and Fig.~\ref{fig.3}(a). The corresponding fluidization times, extracted from the second inflection point of  $\gp(t)$, are shown in Fig.~\ref{fig.4}. First, for the four concentrations explored, we observe a power-law similar to the one found above in the Couette geometry. The exponent is an increasing function of the carbopol weight fraction $C$ (see table~\ref{tab.1}). Second, we checked on a sample of concentration $C=1$~\% wt. that the power law does not depend significantly on the gap width ($1<e<3$~mm) or on the BC. Moreover, since the roughness of the plate surfaces was changed either by using glass plates or by gluing sandpaper, we also infer that the chemical properties of the surfaces do not play any noticeable role in the fluidization process. Thus, we conclude that for a given batch, the power-law behaviour of $\tau_f$ vs $\sigma$ does not depend on the gap width or on the BC.

\section{Discussion}

\begin{figure*}[t]\tt
\centering
\includegraphics[width=0.6\linewidth]{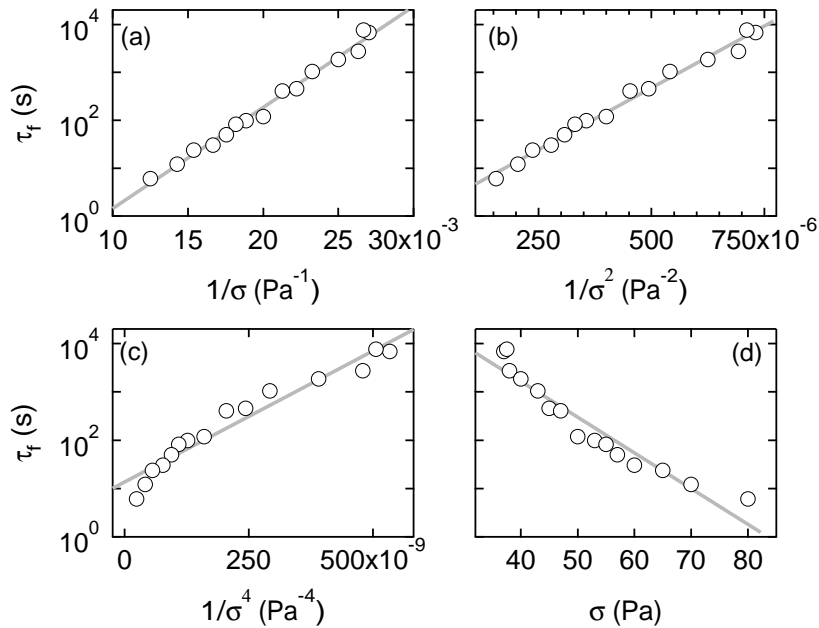}
\caption{Same data set as in Fig.~\ref{fig.2}(b) plotted in various ways: (a) Fluidization time $\tau_f$ vs the inverse of the shear stress $\sigma$. The grey line corresponds to the best linear fit in a log-lin plot corresponding to $\tau_f=\tau_0\exp(\sigma_0/\sigma)$ with $\tau_0=11$ ms and $\sigma_0=485$ Pa; (b) Fluidization time $\tau_f$ vs the inverse of the shear stress squared $\sigma^2$. The grey line corresponds to the best linear fit in a log-lin plot corresponding to  $\tau_f=\tau_0\exp\left[(\sigma_0/\sigma)^2\right]$ with $\tau_0=1$ s and $\sigma_0=110$ Pa;  (c) Fluidization time $\tau_f$ vs $1/\sigma^4$ and (d)  $\tau_f$ vs $\sigma$. The last two semilogarithmic plots allow us to rule out the corresponding behaviours, $\tau_f =\tau_0\exp\left[(\sigma_0/\sigma)^4\right]$ and $\tau_f =\tau_0\exp(-\sigma/\sigma_0)$ respectively.}
\label{fig.5}
\end{figure*} 

 \subsection{Ruling out any fracture-like scenario}

In light of previous works, let us first justify our choice of a power law to fit the $\tau_f$ vs $\sigma$ data. Indeed various other scalings have been proposed in the literature to account for stress-induced fluidization. Analogies with fracture in elastic solids \cite{Caton:2008} lead to $\tau_f =\tau_0\exp\left[(\sigma_0/\sigma)^m\right]$, with $m=1$ for time-dependent rupture in glasses and $m=2$ ($m=4$ resp.) for nucleation of critical cracks that follow the Griffith's criterion in 2D (3D resp.) geometries \cite{Vanel:2009}. Activated processes have also been invoked to support an exponential scaling $\tau_f =\tau_0\exp(-\sigma/\sigma_0)$ found in colloidal gels made of attractive carbon black particles \cite{Gibaud:2010}. Fig.~\ref{fig.5}(c) and \ref{fig.5}(d) clearly show that the last two scalings cannot account for our carbopol data. Still fracture-like processes with $m=1$ or $m=2$ do fit our data correctly as found in \cite{Caton:2008} but the small range of $\sigma$ (less than one decade) does not allow for a definite conclusion derived from rheology alone.

\begin{figure}\tt
\centering
\includegraphics[height=5.4cm]{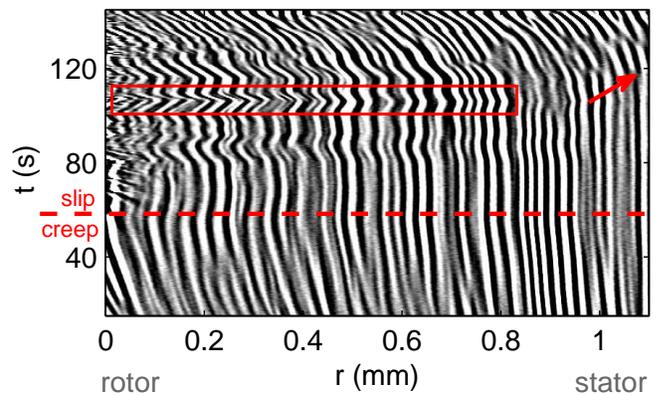}
\caption{Spatiotemporal diagram of the backscattered pressure signals coded in linear gray levels for $\sigma=47$~Pa applied at $t=0$ (batch 1, rough BC). Each horizontal line corresponds to ultrasonic echoes from the seeding glass spheres as a function of the spatial distance $r$ to the rotor for an incident pulse sent at time $t$. Ultrasonic pulses are sent every 0.5~s. The slopes of the echoes in this diagram can be directly interpreted in terms of velocity, a vertical stripe corresponding to a vanishing local velocity.}
\label{fig.6}
\end{figure}

Here, however, one can go one step further since ultrasonic echography provides crucial clues on the local yielding dynamics that rule out a fracture scenario. Indeed Fig.~\ref{fig.6}, which is typical of all shear stresses in rough BC, shows that the backscattered pressure signals do not present any discontinuity in the bulk from one incident pulse to another. This allows us to exclude the presence of fractures at least on scales above a few microns, in contrast with previous observations on a thixotropic organogel (see for instance Fig.~16 and Fig.~17 in reference\cite{Grondin:2008}). Moreover, at short times ($t\lesssim 60$~s), the slopes of the echoes progressively decrease from the rotor to the stator, which is the signature of the homogeneous creep deformation seen in Fig.~\ref{fig.1}(a). For $t\simeq 60$~s, the whole sample suddenly stops as the material fails at the rotor (see the discontinuity at $r\simeq 0$ for $t\simeq 60$~s) and subsequently undergoes total wall slip with a vanishing local shear rate. We interpret the complex shape of the echoes in the slip regime (see, e.g., the red box in Fig.~\ref{fig.6}) as the consequence of elastic bulk deformations due to the large roughness (60~$\mu$m) of the moving wall. Indeed such erratic motions are not seen with smooth BC. Finally the material also fails at the stator at $t\simeq 120$~s (red arrow in Fig.~\ref{fig.6}) so that it slips at both walls and undergoes a solid-body rotation, which is also typical of the slip regime in smooth BC. We can thus keep in mind that fracture planes similar to those found in solids can be ruled out and that carbopol microgels rather fluidize continuously in a liquid-like fashion through the slow growth of a transient shear band.

  \subsection{What can we learn from the creep deformation regime observed with rough BC?}
  
Let us now recall that at short times under applied shear stress and for rough boundary conditions [Fig.~\ref{fig.1} and Fig.~\ref{fig.2}(a)], the shear rate follows a slow decay characterized by a power law $\dot \gamma(t)\propto~t^{-2/3}$ known as the Andrade's law and often observed in solid materials \cite{Andrade:1910,Nechad:2005}. Such a result may seem to contradict our previous paragraph in which we have dismissed the idea of any fracture-like scenario or solid-like behaviour in the fluidization process of carbopol microgels. Actually, there is no contradiction and we shall explain why in this section.

In crystalline materials, creep deformation originates from the collective dynamics of dislocations \cite{Miguel:2002} which is intermittent at the mesoscopic scale \cite{Csikor:2007} but results in an average slow power-law relaxation of the material at a macroscopic scale widely known as Andrade's creep \cite{Miguel:2008}. The discrete dislocation dynamics model has revealed that sufficient -if not necessary-  ingredients to observe Andrade's scaling law appear to be (i) long range anisotropic interactions and (ii) topological constraints \cite{Miguel:2002,Miguel:2008}. Dealing with (i), carbopol microgels are composed of highly crosslinked molecules and thus certainly present long range interactions. Besides, the fact that we found a fluidization time that scales as a power law of the viscous stress could be a signature of such long range interactions, as suggested by numerical results based on fiber bundle models \cite{Kun:2003}. Concerning (ii), carbopol microgels are amorphous materials and thus do not present any dislocations with well-defined topological constraints. Still, since carbopol microgels are made of soft jammed particles linked to each other by polymers, some topological constraints could rather apply to groups of particles presenting a low connectivity, so that both ingredients (i) and (ii) may be present in our soft system.

A last relevant question remains the potential link between the primary Andrade creep regime and the power-law scaling of the fluidization time: does this creep regime govern the behaviour of $\tau_f$ or not? For rough BC, we observed such an initial creep regime and a fluidization time decreasing as a power law of the viscous stress $\sigma-\sigma_c$ (Fig.~\ref{fig.2}). For smooth BC on the other hand, there is no Andrade creep regime and the microgel enters immediately a total wall slip regime [Fig.~\ref{fig.3}(a)]. However we still observe a fluidization time that decreases as a power law of the viscous stress [Fig.~\ref{fig.3}(b)]. Furthermore, the exponent does not seem to depend on the boundary conditions (Fig.~\ref{fig.4}). Finally, it can be checked in Fig.~\ref{fig.1} and Fig.~\ref{fig.2}(a) that the duration of the Andrade creep regime always remains at least one order of magnitude smaller than the fluidization time. Altogether, these findings prove that for carbopol microgels, the Andrade creep regime certainly plays a negligible role on the fluidization power law, contrary to what is seen for fiber composite materials where the primary creep regime is linked to the failure time of the samples \cite{Nechad:2005}. This last point strongly distinguishes the fluidization of carbopol microgels from the failure process of solid materials.

  \subsection{Comparison with experiments under controlled shear rate}

Under imposed shear rate, the fluidization scenario of a carbopol microgel is as follows \cite{Divoux:2010,Divoux:2011}: for rough boundary conditions and in a Couette cell geometry, the microgel first deforms elastically then undergoes plastic deformation and breaks at the rotor before entering a short regime of total wall slip. Then a shear band nucleates at the rotor and slowly invades the whole gap until a homogeneous steady state is reached, with a linear velocity profile. The fluidization time of the carbopol, defined as the time to observe a linear velocity profile, scales as a power law of the applied shear rate: $\tau_f \propto \dot \gamma^{-\alpha}$. It is crucial to emphasize that the whole fluidization sequence is the same under controlled shear rate and under controlled shear stress and that in both cases the fluidization time decreases as a power law of the applied variable. The only difference between these two fluidization processes lies in the duration of the total wall slip regime, which is much shorter in the case of applied shear rate experiments. In the following, we make a connection between the two fluidization dynamics and the steady-state behaviour.

\begin{figure*}[t]\tt
\centering
\includegraphics[height=9cm]{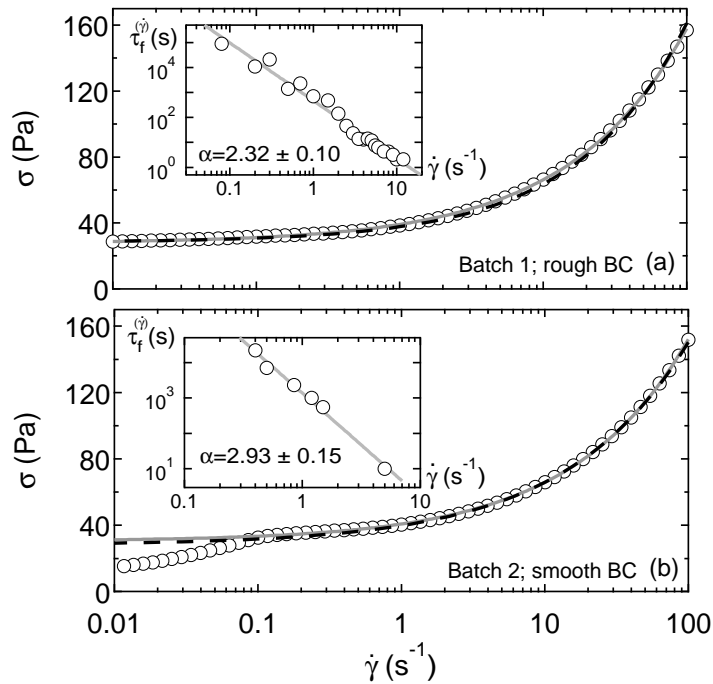}
\caption{(a) Flow curve $\sigma$ vs $\gp$ obtained by decreasing $\gp$ from 100 to 0.01~s with a waiting time of 30~s per point for batch 1 and rough BC. The gray line is the best fit by the HB model [Eq.~\eqref{eq1}] with $\sigma_c=27.8$~Pa, $n=0.53$, and $\tilde{\eta}=11.3$~Pa$\cdot$s$^{-n}$. The black dashed line is the HB model derived from fluidization times with $\sigma_c=28.3$~Pa, $n=0.57$, and $\tilde{\eta}=9.4$~Pa$\cdot$s$^{-n}$ (see text). Inset: fluidization time $\tau_f^{(\gp)}$ vs $\gp$ in controlled shear rate experiments for batch 1 in rough BC, from Divoux {\it et al.}~\cite{Divoux:2010}. The line is $\tau_f^{(\gp)}=A/\gp ^\alpha$ with $A=472\pm 11$ and $\alpha=2.30\pm 0.10$. (b) Same as (a) for batch 2 and smooth BC. The best HB fit (gray line) performed on $0.2 <\gp < 100$~s$^{-1}$ yields $\sigma_c=30.4$~Pa, $n=0.53$, and $\tilde{\eta}=10.3$~Pa$\cdot$s$^{-n}$ and fluidization times (black dashed line) lead to $\sigma_c=28.1$~Pa, $n=0.51$, and $\tilde{\eta}=11.7$~Pa$\cdot$s$^{-n}$. The shape of this flow curve for $\gp<0.1$~s$^{-1}$ is typical of slip phenomena \cite{Meeker:2004a,Meeker:2004b,Ballesta:2008b}. Inset: for batch 2 in smooth BC with $e=0.5$~mm, the best power-law fit of $\tau_f^{(\gp)}$ vs $\gp$ yields $A=1375\pm 46$ and $\alpha =2.93 \pm 0.15$.}
\label{fig.7}
\end{figure*}

 \subsection{Link between transient and steady-state rheology}
 
 \subsubsection{From transient to steady-state rheology.~~} \label{link}
The use of the power law $\tau_f^{(\sigma)} \equiv \tau_f= B/(\sigma-\sigma_c)^{\beta}$ to describe stress-induced fluidization allows for a remarkable connection with independent experiments under imposed shear rate and global steady-state rheology. Indeed, as recalled above, controlled shear rate fluidization on the same carbopol microgels was also shown to involve transient shear banding governed by a power-law behaviour: $\tau_f^{(\gp)} = A/\gp^{\alpha}$, where $\alpha$ only depends on the batch preparation \cite{Divoux:2010}. Since $\tau_f^{(\gp)}$ and $\tau_f^{(\sigma)}$ characterize the same physical process, we may crudely, but rather naturally, write that the two fluidization timescales are proportional:
\begin{equation}
\tau_f^{(\sigma)}=\lambda \tau_f^{(\dot\gamma)},
\end{equation}
where $\lambda$ is a dimensionless constant. This proportionality directly leads to:
\begin{equation}
\sigma=\sigma_c+\tilde{\eta}\,\gp^n\,,
\end{equation}
with $n=\alpha/\beta$ and $\tilde{\eta}=(B/\lambda A)^{1/\beta}$. In other words the HB behaviour that characterizes the {\it steady-state} rheology of our simple YSF is recovered from the two power-law behaviours of the fluidization times in independent {\it transient} experiments. Let us first test this result quantitatively before trying to build it on more solid ground.

\subsubsection{Quantitative test.~~}
This link between local fluidization dynamics and global rheology is tested quantitatively in Fig.~\ref{fig.7} for batches 1 and 2 and different BC. Going back to the $\tau_f^{(\gp)}$ vs $\gp$ data (insets of Fig.~\ref{fig.7}), we predict $n=2.3/4.0=0.57$ for batch 1 and rough BC and $n=2.93/5.75=0.51$ for batch 2 and smooth BC. These exponents are in very good agreement with those extracted from the best HB fits of the flow curves, $n=0.53$. It is also quite remarkable that the values of $\sigma_c$ inferred from our fitting procedure of $\tau_f^{(\sigma)}$ ($\sigma_c=28.3$ and 28.1~Pa, Fig.~\ref{fig.2}) coincide with those of the best HB fits ($\sigma_c=27.8$ and 30.4~Pa) to within 10~\%. As a consequence the HB models inferred from the fluidization times are undistinguishable from the best HB fits of the steady-state rheology provided $\lambda$ is left as a free parameter to achieve the correct values of $\tilde{\eta}$ from the prefactors $A$ and $B$ of the two power laws. 
 We get $\lambda = 10.2$ for batch 1 and 9.1 for batch 2, which means that fluidization under a given $\sigma$ is roughly ten times slower than fluidization under the corresponding $\gp$. This difference most probably results from the fact that our YSF undergoes a long-lasting slip regime under stress while the slip regime is much shorter under imposed shear rate.

\begin{figure*}[t]\tt
\centering
\includegraphics[width=15cm]{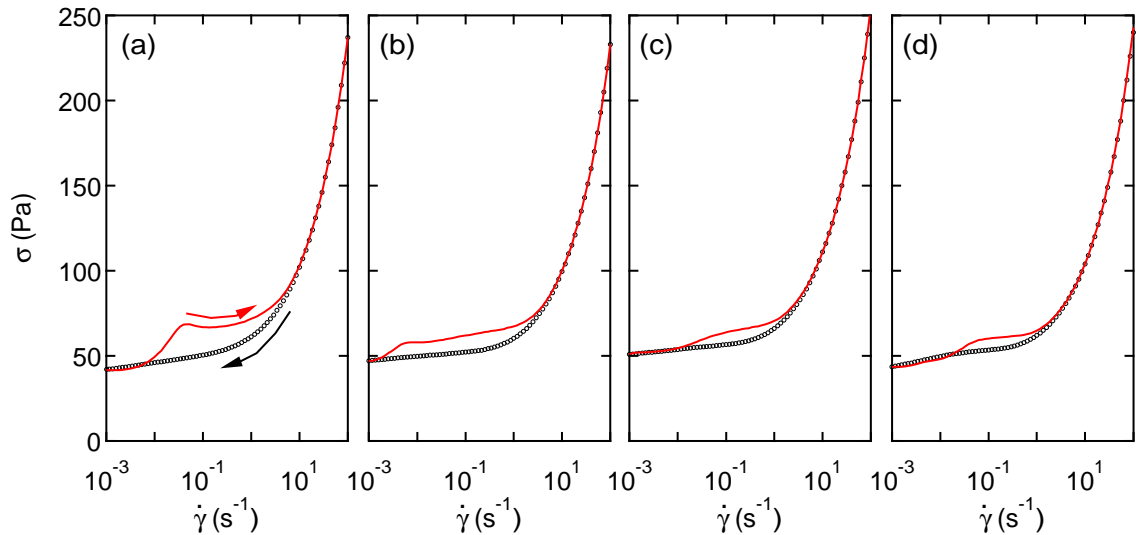}
\caption{Flow curves, shear stress $\sigma$ vs shear rate $\dot \gamma $, obtained by decreasing $\dot \gamma$ from 100 to $10^{-3}$~s$^{-1}$ ($\circ$) and then increasing $\dot \gamma$ from $10^{-3}$ to 100~s$^{-1}$ (red line) for various waiting times per point: (a) $t_w=2$~s, (b) $t_w=10$~s, (c) $t_w=30$~s, and (d) $t_w=70$~s. Note that the hysteresis loop for $\dot \gamma \lesssim 1$~s$^{-1}$ decreases for increasing waiting times. The total duration of the longest measurements shown in (d) is $2.1\cdot 10^{4}$~s. Experiments performed in a plate-plate geometry ($e=1$~mm) with rough BC ($\delta=46~\mu$m) on a 1~\% wt. carbopol microgel without any seeding glass spheres.}  
\label{fig.8}
\end{figure*} 

\subsubsection{Discussion.~~}
Let us now further justify writing that the two fluidization timescales $\tau_f^{(\gp)}$ and $\tau_f^{(\sigma)}$ are proportional. First, we emphasize once again that both fluidization processes under imposed shear rate and under imposed shear stress are rigorously the same in that they present the exact same sequence of local events. Second, although the first two steps of the fluidization process, namely the homogeneous strain and the total wall slip regimes, present a shorter duration in the case of applied shear rate, the fluidization times are of the same order of magnitude for both applied shear rate and applied shear stress. Third, a proportionality law is the simplest link one can think of between two timescales without artificially invoking another characteristic time whose origin and interpretation would be unclear. The fact that we can make any additional timescale irrelevant will be confirmed in the next section.

Now, our interpretation of the HB exponent as the ratio of two fluidization exponents allows us to revisit some recent observations. Indeed, quite similar results have been observed on acidic solutions of type I collagen by Gobeaux {\it et al.}~\cite{Gobeaux:2010}. On the one hand, the steady-state rheology of type I collagen molecules is well described by a power law  $\sigma = K\dot \gamma ^n$, with $n \simeq 0.27$, i.e. a HB model with no yield stress. On the other hand, creep experiments have revealed that the fluidization occurs after a lag time $t_c$ that scales as $t_c \propto 1/\sigma^{1/\alpha}$. This can also be read as $t_c \propto 1/(\sigma-\sigma_c)^{1/\alpha}$ with $\sigma_c=0$. Using an original time stress superposition principle, the authors have shown that $\alpha$ and $n$ are equal within error bars. Here, our approach allows us to predict that acidic solution of the same type I collagen should also fluidize under an applied shear rate $\dot \gamma$ after a lag time $t_c$ which should scale as $t_c \propto 1/\dot \gamma^\beta$ with $\beta=1$, so that if one imposes the lag time for both fluidization processes to be propotional, one would recover the steady-state rheology with $n=\alpha$. Such a prediction remains to be experimentally verified and could reinforce our claim that the two fluidization timescales are proportional.

Last but not least, in a recent work on bidimensional wet foams, Katgert {\it et al.}~\cite{Katgert:2009} also proposed an interpretation of the HB exponent $n$. The authors made a connection between the way the average drag force on a bubble scales with the velocity and the power-law behaviour of the viscous stress $\sigma-\sigma_c$  in the HB model. In this framework, $n$ appears as a direct measure of the average forces at the bulk level taking into account both the local interbubble drag and the disorder induced by the flow. This is also compatible with our results, as we have observed that the fluidization exponents $\alpha$ and $\beta$ are sensitive to the size distribution and/or the spatial organization of the soft particles that constitute the microgel, through the preparation protocol and the carbopol concentration (Fig.~\ref{fig.4}).

 \subsection{Time-dependent effects and hysteresis cycles}

In this subsection, we provide the reader with further evidence showing that it would be artificial to invoke any other characteristic time or any hidden dynamical variable besides $\tau_f$ to describe the fluidization process of carbopol microgels. This point is strongly related to the fact that these microgels are {\it simple} YSF.

Thixotropic and simple YSF are usually placed in two categories which preclude one another \cite{Moller:2009b}. If satisfying at first sight, such a rough description remains qualitative and has resulted in defining useless sub-categories such as ``unusual yield stress fluids" \cite{Denn:2010}. In order to overcome those difficulties, Coussot and Ovarlez \cite{Coussot:2010} recently proposed an interesting reunification of these two categories within a single theoretical framework by introducing the ratio $D$ of two timescales: a characteristic relaxation time $\eta_0/G_0$ (built on the viscosity $\eta_0$ and the elastic modulus $G_0$ of the fluid) and the restructuration time $\theta$ of the system. A fluid with $D=\eta_0/(G_0 \theta)$ close to 1 would be a simple YSF for which the restructuration time is indeed roughly equal to the relaxation time, whereas a fluid with $D\ll 1$ would correspond to a thixotropic material, concomitantly presenting aging effects and restructuration over long durations. The key point of such a description is that one goes continuously from one type of fluid to the other simply by tuning the timescale ratio $D$. 

A simple way to probe the relevant timescales consists in performing successive decreasing and increasing ramps of controlled shear rates. Fixing the shear rate range (here, $\dot \gamma_{\rm min}=10^{-3}<\dot \gamma<\dot \gamma_{\rm max}=10^2$~s$^{-1}$) and the number of experimental data points (here, 15 points per decade), the only control parameter is the {\it waiting time per point} $t_w$ spent at each imposed value of $\dot \gamma$. In other words, $t_w^{-1}$ is the rate at which we scan the flow curve $\sigma(\dot \gamma)$.  In Fig.~\ref{fig.8}, we report flow curves obtained with four different values of $t_w$ on a 1 \% wt. carbopol microgel. As already briefly mentioned in \cite{Divoux:2010}, for a given value of $t_w$, we observe a slight hysteresis between decreasing and increasing shear-rate sweeps. Let us emphasize here that such an effect is also noticeable but not discussed in previous literature \cite{Coussot:2009,Moller:2009a} and that it is thus not particular to the type of carbopol that we are using.

\begin{figure}[t]
\centering
\includegraphics[width=0.9\columnwidth]{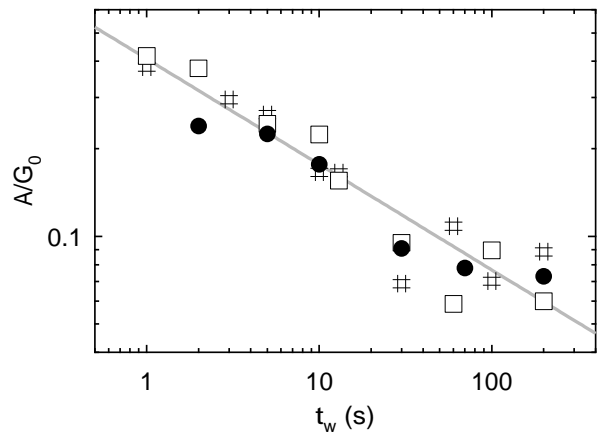}
\caption{Area ${\cal{A}}$ of the hysteresis between the decreasing and the increasing flow curve vs the waiting time per point $t_w$ for various carbopol weight fractions (symbol, \% wt. carbopol): ($\bullet$, 1 \%); ($\sharp$, 2 \%); ($\square$, 3 \%). ${\cal{A}}$ decreases as a power law of the waiting time: ${\cal{A}} /G_0= 0.41/t_w^{0.36}$, where $G_0$ is the elastic modulus of the microgel. Data obtained in a plate-plate geometry ($e=1$~mm) with rough BC ($\delta=46~\mu$m) on carbopol microgels without any seeding glass spheres.}  
\label{fig.9}
\end{figure}

Repeating the shear rate sweep for different waiting times, we observe that the area ${\cal{A}}$ of the hysteresis loop, defined as 
\begin{equation}
{\cal{A}} \equiv \int_{\log \dot \gamma_{\rm min}}^{\log \dot \gamma_{\rm max}} \sigma[\log(\dot \gamma~')]d[\log(\dot \gamma~')],
\end{equation} 
decreases as power law of $t_w$ with an exponent 0.36 [Fig.~\ref{fig.9}]. First note that this result is robust: the same trend is observed for carbopol microgels of three different mass concentrations ($C=1$, 2, and 3 \%) and, when rescaled by the elastic modulus $G_0$ of the microgel, ${\cal{A}}$ is roughly independent of $C$. Second, in light of the work by Coussot and Ovarlez \cite{Coussot:2010} detailed above, this result is (i) compatible with a simple YSF, ruling out any thixotropic behaviour, and (ii) suggests that no other timescale than the fluidization time $\tau_f$ is necessary to describe the flow behaviour of carbopol microgels. Indeed, in the case of a simple YSF, for $t_w \gtrsim 1$~s, $t_w$ is generally large enough compared to the characteristic relaxation time of the fluid so that $t_w \gg \eta_0/G_0 \sim \theta$. Thus, the larger the imposed value of $t_w$, the less the effects of the fluid relaxation and restructuration will be probed and so the smaller the area of the hysteresis loop. In other words, the larger $t_w$, the more the carbopol microgel ``forgets" about its shear history. 


In the case of a thixotropic YSF, the restructuration timescale would be of hundreds of seconds or more so that $\eta_0/G_0 \ll \theta$ and $t_w$ would be of the same order of magnitude as $\theta$. Thus, in the thixotropic case, at least two timescales, $\theta$ and $t_w$, would be involved in the material dynamics during shear rate sweeps and one would expect a more complex behaviour of ${\cal{A}}$ vs $t_w$ than a simple decreasing function. In particular, it is anticipated that the hysteresis loop grows larger as long as $t_w<\theta$ and that it decreases (or at least saturates) for $t_w\gg\theta$ when the restructuration timescale is no longer relevant. An extensive study of the behaviour of hysteresis cycles for both simple and thixotropic YSF is currently underway to deeper test these ideas. In any case, to us, the results shown in Fig.~\ref{fig.9} provide a strong confirmation that (i) carbopol microgels are simple YSF which exhibit negligible hysteresis when $t_w$ is large enough and (ii) no other timescale or hidden parameter is needed to describe the microgel rheology as the hysteresis can be accounted for only by a transient shear banding phenomenon which is fully described by the timescale $\tau_f$. 

\section{Summary, open questions, and outlook}

\subsection{Summary}

We have performed a temporally and spatially resolved study of the stress-induced fluidization of a simple yield stress fluid. The fluidization is a four step process that successively involves Andrade-like creep deformation, a total wall slip regime, and a transient shear banding phenomenon that leads to a homogeneous flow in steady state. The time to reach a linear velocity profile is a robust decreasing power law of the applied shear stress which neither depends on the boundary conditions nor on the gap width, while the exponent is a function of the microgel microstructure. One of the key results of this article is that the exponent $n$ in the HB model which describes the steady-state rheology naturally appears as the ratio $\alpha/\beta$ of two fluidization exponents derived from independent experiments under controlled stress and under controlled shear rate. To our knowledge, this provides for the first time a clear link between the {\it transient regime} of the fluidization process and the {\it steady-state} rheology. 

\subsection{Open questions and outlook}

We would like to speculate that this last result is general for simple YSF and future experiments will focus on measuring $\tau_f^{(\sigma)}$ and $\tau_f^{(\gp)}$ in emulsions and wet foams so as to extract the value of the exponent $\alpha$ and $\beta$ and test their link with the steady-state rheology. Concerning Carbopol microgels, it would also be of valuable interest to unambiguously link the microscopic properties of the microgel, in particular the size of the microstructure, to the value of the fluidization exponents.

The transient shear banding scenario, common to both applied shear stress and shear rate experiments, also remains to be characterized at a microscopic scale. In the case of traditional steady-state shear banding, the two flowing bands present two different microstructures \cite{Lerouge:2010}. In the case of wormlike micelle solutions for instance, the highly sheared band presents a nematic-like order whereas the micelles are more entangled in the weakly sheared band. Here, for the transient shear banding observed during the fluidization of carbopol microgels, one may wonder if there is any structural difference between the flowing band and the arrested region.

Another puzzling issue comes up when one compares the fluidization laws of two different soft systems: carbopol microgels whose fluidization time decreases as a power law of the viscous stress, and weakly attractive carbon black gels whose fluidization time decreases exponentially with the applied stress \cite{Gibaud:2010}. While the latter system is a fractal colloidal gel with a low volume fraction, carbopol microgels are constituted of jammed swollen particles. How and why does such a structural difference lead to different stress-induced fluidization law? Could one tune continuously the system properties to switch from one fluidization behaviour to the other?

Finally, we wish to emphasize that it would be very interesting to compare the present experimental data on stress-induced fluidization to theoretical predictions. Unfortunately, to the best of our knowledge, most recent theoretical works on shear banding in yield stress materials have focused on stationary states only \cite{Moller:2008,Coussot:2010}. One may think of using standard models for time-dependent materials based on structure-dependent kinetic equations \cite{Mewis:2009,Coussot:2002a,Coussot:2007} but, as discussed above, the restructuration kinetics does not appear as a relevant ingredient for simple YSF such as carbopol microgels. Two recent theoretical papers \cite{Fielding:2009,Moorcroft:2011} based on the soft glassy rheology (SGR) model do address the transient regime but a detailed study of the fluidization times remains to be performed. Therefore, no analytical or numerical prediction is yet available for the fluidization behaviour observed in the present experimental work. Such a prediction would constitute a major step towards a full understanding of the yielding dynamics in simple YSF.

\begin{acknowledgments}
We thank Y. Forterre for providing us with the carbopol, D.~Tamarii and V.~Grenard for substantial help with the experiments and with the software, and H.~Feret for technical help. We also thank L.~Bocquet, A.~Colin, and S.~Santucci for several enlightening discussions.
\end{acknowledgments}

\bibliography{bibdivoux} 
\bibliographystyle{rsc} 

\end{document}